\documentclass[aps,prd,12point,twocolumn,nofootinbib,showpacs,superscriptaddress]{revtex4-1}
\def\theequation{\arabic{section}.\arabic{equation}}
\usepackage{psfrag}
\usepackage{subfigure}
\usepackage{color}
\usepackage{mathrsfs}
\usepackage{graphicx}
\usepackage{amssymb, bm}
\usepackage{amsmath, amsthm}
\usepackage{epstopdf}
\usepackage{hyperref}
\usepackage{enumerate}
\usepackage{longtable}
\usepackage{amsfonts}
\usepackage[T1]{fontenc}
\usepackage{bm}
\usepackage{amsmath}  
\usepackage{graphicx} 
\usepackage{color}
\usepackage{appendix}
\usepackage{amsmath}

\DeclareMathOperator{\arcsinh}{arcsinh}
\DeclareMathOperator{\arctanh}{arctanh}

\newcommand{\be}{ \begin{equation}}
\newcommand{\ee}{\end{equation}} 
\begin{document}
\def\theequation{\arabic{section}.\arabic{equation}} 

\title{A new symmetry of the spatially flat Einstein-Friedmann equations}

\author{Steve Dussault}
\email[]{sdussault19@ubishops.ca}
\affiliation{Department of Physics and Astronomy, Bishop's University, 
2600 College Street, Sherbrooke, Qu\'ebec, 
Canada J1M~1Z7}

\author{Valerio Faraoni}
\email[]{vfaraoni@ubishops.ca}
\affiliation{Department of Physics and Astronomy, Bishop's University, 
2600 College Street, Sherbrooke, Qu\'ebec, 
Canada J1M~1Z7}



\begin{abstract}

We report a new symmetry of the Einstein-Friedmann equations for spatially 
flat Friedmann-Lema\^itre-Robertson-Walker universes. We discuss its 
application to barotropic perfect fluids and its use as a 
solution-generating technique for scalar field universes.

\end{abstract}


\maketitle

\section{Introduction}
\setcounter{equation}{0}
\label{sec:1}

In general-relativistic cosmology, the matter content of the universe is 
typically modelled by 
a perfect fluid with energy density $\rho(t)$ and isotropic pressure $P(t)$ 
related by a barotropic equation of state $P=P(\rho)$. This 
fluid is described by the stress-energy 
tensor 
\be
T_{ab}= \left(P+\rho \right) u_a u_b +P g_{ab} \,,
\ee
where $u^c$ is the fluid four-velocity, which coincides with 
the time direction 
of comoving observers (here we follow the notations of 
Ref.~\cite{Wald}, using units in which the speed of light is unity).

Assuming spatial homogeneity and isotropy, the 
geometry is necessarily the Friedmann–Lema\^itre–Robertson–Walker (FLRW) 
one, with line element 
\be
ds^2 = g_{ab} dx^a dx^b= -dt^2 +a^2(t) \left( \frac{dr^2}{1-Kr^2}+r^2 
d\Omega_{(2)}^2 \right) 
\ee
in comoving coordinates $\left( t, r, \vartheta, 
\varphi \right)$, 
where $g_{ab}$ is the metric tensor, $a(t)$ is the cosmic scale factor, 
$K$ is the curvature index,  and 
$d\Omega_{(2)}^2=d\vartheta^2 
+\sin^2 \vartheta \, d\varphi^2$ is the line element on the unit 2-sphere 
\cite{Wald}. In this geometry, the Einstein field equations 
\begin{eqnarray}
R_{ab} - \frac{1}{2} \, g_{ab} R  + \Lambda  g_{ab} = 
 {8 \pi G } T_{ab}
\end{eqnarray}
(where $R_{ab} $ is the Ricci tensor, $R \equiv g^{ab}R_{ab}$, and 
$\Lambda$ is the cosmological constant) reduce 
to the Einstein-Friedmann equations of relativistic 
cosmology
\begin{eqnarray}
&&H^2 \equiv \left( \frac{\dot{a}}{a}\right)^2 
= \frac{8\pi G}{3} \, \rho +\frac{\Lambda}{3} 
-\frac{K}{a^2}\,, \label{eq:1}\\
&&\nonumber\\
&& \dot{\rho}+3H\left(P+\rho \right)=0\,, \label{eq:2}\\
&&\nonumber\\
&&\frac{\ddot{a}}{a}= -\, \frac{4\pi G}{3} \left( \rho +3P \right) 
+\frac{\Lambda}{3} \,, \label{eq:3}
\end{eqnarray}
where an overdot denotes 
differentiation with respect to $t$, $H(t)\equiv \dot{a}/a$ is the 
Hubble function, and the constant $K$ describes a closed universe if 
it is positive, a 
spatially flat one if it vanishes, or a hyperbolic one if $K<0$. 

Out of the three Einstein-Friedmann equations~(\ref{eq:7})-(\ref{eq:9}), 
only two are independent. If any two are given, the third one can be 
derived from them. Without loss of generality, we take the Friedmann 
equation~(\ref{eq:1}) and the energy conservation equation~(\ref{eq:2}) as 
primary, with the acceleration equation~(\ref{eq:3}) following from them.

Symmetries are important for any physical theory and, naturally, there 
is a wealth of literature on the symmetries of the Einstein-Friedmann 
equations. Some of these studies are inspired by string dualities, 
although they are not always directly related to them 
\cite{Chimento02,SD2,SD3,SD4,SD5,SD6,SD7,SD8,SD9,SD10, 
SD11,SD12,SD13,SD14,SD15,SD16,SD17,SD18,SD19,SD20,SD21};  others are 
based on methods of supersymmetric quantum mechanics 
\cite{susy1,susy2,susy3}, and other times FLRW symmetries are 
studied in relation with solution-generating techniques ({\em e.g.}, 
\cite{SD15,SD17,Chimento02,Pailasetal20,Lieolder,Borowiec14,Mariafelicia15, 
Mariafelicia16,Mariafelicia20}). Here we propose a new symmetry of these 
equations and discuss its possible uses for FLRW universes filled with  a 
perfect fluid with barotropic, linear, and constant equation of state and 
for scalar fields minimally coupled to the spacetime curvature. This 
symmetry generalizes one previously introduced in Ref.~\cite{Chimento02} 
(see also \cite{SD2,SD3,SD5,SD9,SD10,SD13,SD14,SD20}) and studied recently 
in Ref.~\cite{Pailasetal20}.

\section{A new symmetry of the Einstein-Friedmann equations}
\label{sec:2}
\setcounter{equation}{0}

The new symmetry transformation involves a rescaling of the comoving time 
$t$ (and, consequently, of the Hubble function $H$ and of all first time 
derivatives), and the fluid energy density and pressure. Since, in 
general, the 
pressure is rescaled differently than the energy density, the equation of 
state changes under the symmetry transformation. The Einstein-Friedmann 
equations remain invariant in form. This symmetry only applies to 
spatially flat universes and is 
given by
\begin{eqnarray}
dt & \rightarrow &  d\bar{t}=f(\rho)dt \,, \label{eq:7}\\
&&\nonumber\\ 
\rho & \rightarrow & \bar{\rho}=\frac{\left(1-f^2\right)}{8\pi G f^2} 
\Lambda + \frac{\rho}{f^2}  
= \frac{(1-f^2)}{f^2} \, \rho_{\Lambda} +\frac{\rho}{f^2} \,, 
\nonumber\\
&&\label{eq:8}\\
&& \nonumber\\
P & \rightarrow &  \bar{P }=-\bar{\rho}+ 
\frac{\left[4\pi G f-\left(8\pi G \rho+\Lambda\right) f'\right]}{ 
4\pi G f^3} \left(P+\rho\right)  \nonumber\\
&&\nonumber\\
&\, & = -\bar{\rho} + \frac{ \left[ f-2\left( \rho+\rho_{\Lambda} \right) 
f'\right]}{f^3} \left(P+\rho\right) \,, \label{eq:9}
\end{eqnarray}
where $f(\rho) $ is a regular, positive, and dimensionless function, $f'$ 
is its derivative, and 
$\rho_{\Lambda} \equiv \Lambda/\left(8\pi G\right)$ is the effective 
energy density 
of the 
cosmological constant.  
Since $\rho=\rho(t)$, the new differential $d\bar{t}=f(\rho(t)) dt$ is 
exact, with $ \tilde{t}(t)= \int dt f(\rho(t)) $. 
The inverse transformation is 
\begin{eqnarray}
dt&=& \frac{d\bar{t}}{f} \,, \label{1inverse}\\
&&\nonumber\\
\rho &=& f^2 \bar{\rho} + \left(f^2-1\right) \rho_{\Lambda} 
\,, \label{2inverse} \\
&&\nonumber\\
P+\rho &=& \frac{f^3}{ f- 2\left( \rho + \rho_{\Lambda} \right) 
f'} \left(\bar{P}+\bar{\rho} \right) \,. \label{3inverse}
\end{eqnarray}
Using 
\begin{eqnarray}
\dot{a} &\equiv& \frac{da}{dt}= \frac{da}{d\bar{t}}\, 
\frac{d\bar{t}}{dt} =f \, \frac{da}{d\bar{t}} \,, \label{eq:adot}\\
&&\nonumber\\
H&=& f \bar{H} \equiv \frac{f}{a} \, \frac{da}{d\bar{t}} \,,
\end{eqnarray} 
and Eq.~(\ref{2inverse}), the Friedmann equation~(\ref{eq:1}) with $K=0$ 
yields 
\be
f^2 \bar{H}^2 = \frac{8\pi G}{3} \left( f^2 \, \bar{\rho}+\frac{f^2}{8\pi 
G} \, \Lambda \right)\,
\ee
and, finally, 
\be
\bar{H}^2 = \frac{8\pi G}{3} \, \bar{\rho} +\frac{\Lambda}{3} \,.
\label{Friedmanneqbarred}
\ee
Let us verify the covariant conservation equation in the barred variables.  
We have
\begin{eqnarray}
\frac{d\bar{\rho}}{d\bar{t}} &=&  
\frac{1}{f} \, \frac{d}{d\bar{t}} \left[ 
\frac{ (1-f^2) }{ f^2}\, \rho_{\Lambda} +\frac{\rho}{f^2} \right] 
\nonumber\\
&&\nonumber\\
&=& \frac{  f - 2\left(\rho + \rho_{\Lambda} \right) f'}{ f^4} 
\, \dot{\rho} \,,
\end{eqnarray}
  so that
\begin{eqnarray}
&& \frac{d\bar{\rho}}{d\bar{t}} +3\bar{H} \left( \bar{P}+\bar{\rho} 
\right) 
\nonumber\\
&&\nonumber\\
&& = \left[ \frac{ f- 2\left( \rho + \rho_{\Lambda} \right) f'}{f^4} 
\right] 
\left[ \dot{\rho}+3H \left(P+\rho\right) \right] =0 \nonumber\\
&& \label{conservationbarred}
\end{eqnarray}
by virtue of the ``old'' conservation equation~(\ref{eq:2}). The 
acceleration equation in barred variables is automatically satisfied since 
it can be easily derived from the other two Einstein-Friedmann 
equations~(\ref{Friedmanneqbarred}) and~(\ref{conservationbarred}).

The symmetry (\ref{eq:7})-(\ref{eq:9}) is not contained in those discussed 
in Refs. \cite{SD21,VFSymmetry2020}. When $\Lambda=0$, 
Eq.~(\ref{eq:8}) simplifies considerably and gives
\be
f(\rho)=\sqrt{ \frac{\rho}{\bar{\rho}} } \,,
\ee
and then the symmetry (\ref{eq:7})-(\ref{eq:9}) reduces to the one found 
by Chimento \cite{Chimento02} and studied recently in \cite{Pailasetal20}.  In fact, this 
transformation consists of  \cite{Chimento02} 
\begin{eqnarray}
 \rho &\rightarrow &\bar{\rho}(\rho) \,,\label{chimento1}\\
&&\nonumber\\
H &\rightarrow &\bar{H}= \sqrt{ \frac{ \bar{\rho}}{\rho}} \, H \,, 
\label{chimento2}\\
&&\nonumber\\
P+\rho &\rightarrow &\bar{P}+\bar{\rho}= \left(P+\rho \right) \sqrt{ 
\frac{ \rho}{\bar{\rho}} } \, \frac{d\bar{\rho}}{d\rho} 
\,.\label{chimento3}
\end{eqnarray}
The comparison of Eq.~(\ref{chimento1}) and our Eq.~(\ref{eq:8}) with 
$\Lambda=0$ yields $\bar{\rho}(\rho) = \rho/f^2$ and
\be
f(\rho) =\sqrt{ \frac{ \rho}{ \bar{\rho}} }\,,
\ee
using which, Eq.~(\ref{chimento3}) gives our Eq.~(\ref{eq:9}). 
Furthermore, using Eq.~(\ref{eq:adot}), the comoving time  
rescaling~(\ref{eq:7}) gives 
immediately the rescaling~(\ref{chimento2}) of the Hubble function.  
Therefore, our transformation~(\ref{eq:7})-(\ref{eq:9}) constitutes a 
generalization 
of the symmetry (\ref{chimento1})-(\ref{chimento3}) of 
Ref.~\cite{Chimento02} to the case in which a  cosmological 
constant $\Lambda$ is present.

A de Sitter universe fueled by a positive cosmological constant is a  
fixed point 
of the transformation. In fact, if $\Lambda> 0$ and $P=\rho=0$, 
Eq.~(\ref{eq:9}) yields $\bar{P}=-\bar{\rho}=$~const., while 
Eq.~(\ref{eq:8}) gives 
\be
\bar{\rho}=\frac{(1-f^2)}{f^2} \, \rho_{\Lambda} \,;
\ee
however, since $\bar{\rho}$ is constant, it must be $f=$~const. Then, the 
constant $f$ can be absorbed into a change of unit of the comoving time 
$t$  
and set to unity. Hence, the 
transformation~(\ref{eq:7})-(\ref{eq:9}) 
reduces to the identity and maps a de Sitter space into itself.

Explicit examples of transformations mapping perfect fluid universes with 
an equation of state of the form $P=w\rho$, $w=$~const. into FLRW 
universes with a fluid with 
non-linear equations of state and $\Lambda$ are shown in the Appendix.

\section{Perfect fluids with $w=$~const.}
\setcounter{equation}{0}
\label{sec:3}

Perfect fluids with barotropic, linear, and constant equation of state of 
the form $P=w\rho$ with $w=$~const. are of great interest in cosmology. 
Therefore, it is particularly relevant to consider symmetry 
transformations that map one such fluid into another one. Beginning from a 
fluid with equation of state $ P=w\rho $ and writing 
$\bar{P}=\left(\bar{w}+1 \right) \bar{\rho}$,   Eq.~(\ref{eq:9}) gives
\be
\bar{w}+1= \left( w+1\right) \frac{\rho}{\rho + (1-f^2) \rho_{\Lambda}} \, 
\frac{ f-2\left( \rho +\rho_{\Lambda} \right) f'}{f} \,;
\ee
by imposing that $\bar{w} \equiv \bar{P}/\bar{\rho} =$~const. and 
introducing
\be
s \equiv \frac{\bar{w}+1}{w+1} = \mbox{const.} \,,
\ee
one finds an ordinary differential equation that must be satisfied in 
order to take 
a perfect fluid with constant equation of state into another:
\be
f' + \frac{(s-1) f}{2\left( \rho+\rho_{\Lambda} \right)} + s f\left( 1-f^2 
\right) \frac{ 
\rho_{\Lambda}}{2\rho \left( \rho +\rho_{\Lambda} \right)} =0 \,. 
\label{ODE}
\ee
This equation is nonlinear and, in general, it is difficult to find 
analytic solutions $f(\rho)$. However, we can present one. Consider the 
choice
\be
f(\rho) = \sqrt{ \frac{\rho+\rho_{\Lambda}}{ \gamma \rho^{\alpha} 
+\rho_{\Lambda} } } \,,\label{putative}
\ee
where $\alpha$ and $\gamma$ are constants, with $\alpha$ dimensionless and 
$ \left[ \gamma \right] = \left[ \rho^{1-\alpha} \right]$. 

Substituting the putative solution~(\ref{putative}) into the 
first order equation~(\ref{ODE}), straightforward manipulations lead to
\be
\left(s-\alpha\right) \left( \rho+\rho_{\Lambda} \right)=0 \,,
\label{algebraic}
\ee
which is satisfied if  $ \alpha=s$.

If $\rho=\rho_{\Lambda}$, then the perfect fluid is a cosmological 
constant, the solution is de Sitter space,  
$f=\sqrt{\frac{2}{\gamma \, \rho^{\alpha-1}_{\Lambda} +1}}=$~const., 
and $f$ can be absorbed into a redefinition of the unit of time.

Under the transformation~(\ref{eq:7})-(\ref{eq:9}) with 
the choice~(\ref{putative}) of the function $f(\rho)$, the energy density 
and pressure transform as 
\begin{eqnarray}
\rho \rightarrow \bar{\rho}&=& \gamma\, \rho^{\alpha} \,,\\
&&\nonumber\\
P \rightarrow \bar{P}&=& \gamma\left[ \alpha(w+1)-1\right]  
\rho^{\alpha} \,.
\end{eqnarray}
Combining these two equations yields $ \bar{P}=\left[ \alpha (w+1) -1 
\right] \bar{\rho} $ or $\bar{P}= \bar{w} \, \bar{\rho} $ with
\be
\bar{w}= \alpha \left( w+1\right)-1 \,,
\ee
where the constant $\gamma$ disappears from the equation of state.  
Special choices of the constants $w$ and $ \alpha$ include the 
following.\\

\noindent $\bullet$ $w=-1$ implies $\bar{w}=-1 \;\; \forall \alpha$: 
this is again the result that de Sitter space is a fixed point of the 
transformation.\\\\
\noindent $\bullet$ A dust with $w=0$ is mapped into a 
perfect fluid with $\bar{w}=\alpha-1$. If $\alpha=1$, a FLRW space 
with dust is a fixed point and is mapped into another FLRW space filled 
with dust. This is a special case of the more general result below.\\\\
\noindent $\bullet$ If $\alpha=1$, one has $\bar{w}=w$, that is, any fluid 
with $w=$~const. is a fixed point of the transformation with function
\be
f_1 (\rho) =\sqrt{ \frac{\rho+\rho_{\Lambda}}{ \gamma\, 
\rho+\rho_{\Lambda}} } 
\,.
\ee
This result could have been guessed from the fact that the (now dimensionless) 
constant 
$\gamma$ disappears from the barred equation of state, therefore one could 
choose it to be unity, making $f \equiv 1$.\\\\ 
\noindent $\bullet$ If $\alpha=4/3$, then $\bar{w}= (4w+1)/3$ and a dust 
($w=0$) is mapped into radiation ($\bar{w}=1/3$).\\\\
\noindent $\bullet$ If $\alpha=3/2$, a radiation fluid ($w=1/3$) is mapped 
into a stiff fluid ($\bar{w}=1$).

\section{Scalar fields}
\setcounter{equation}{0}
\label{sec:4}

A scalar field $\phi$ minimally coupled to the curvature in a FLRW space 
behaves as an irrotational perfect fluid \cite{Madsen1, 
Madsen2,Pimentel,Faraoni12,Semiz12,Ballesteros,Jeremy}. A choice 
of the scalar field potential $V(\phi)$ corresponds, roughly speaking, to 
a choice of the equation of state for the effective fluid, but this is not 
a one-to-one correspondence \cite{Madsen2, myAJP2, 
BayinCooperstockFaraoni}.

The 
energy density and pressure are
\begin{eqnarray}
\rho_{\phi} &=&\frac{\dot{\phi}^2}{2} + V(\phi) \,, \label{rhophi}\\
&&\nonumber\\
P_{\phi} &=&\frac{\dot{\phi}^2}{2} - V(\phi) \,,\label{Pphi}
\end{eqnarray}
respectively. The 
covariant conservation equation for this effective fluid coincides with 
the Klein-Gordon equation
\be
\ddot{\phi}+3H \dot{\phi}+\frac{dV}{d\phi}=0 \,.\label{KleinGordon}
\ee
It is clear from Eqs.~(\ref{rhophi}) and~(\ref{Pphi}) that a free 
massless scalar field behaves like a stiff fluid with equation of state 
$P=\rho$. We will attempt to generate scalar field solutions with a 
potential $V(\phi)$ in a spatially flat FLRW 
starting from this free field solution. 

We begin by noting that, since $d\phi/d\bar{t} = f^{-1} \, d\phi/dt$,  
after the symmetry 
transformation~(\ref{eq:7})-(\ref{eq:9}), the ``new'' energy density of 
this effective scalar field  fluid is
\begin{eqnarray}
\bar{\rho} &=& 
\frac{1}{2f^2} \left( \frac{d\phi}{dt} \right)^2 +\frac{(1-f^2)}{f^2} \, 
\rho_{\Lambda} \nonumber\\
&&\nonumber\\
&=&  \frac{1}{2} \left( \frac{d\phi}{d\bar{t}} \right)^2 
+\frac{(1-f^2)}{f^2} \, 
\rho_{\Lambda} \,.
\end{eqnarray}
By choosing
\be 
f= \sqrt{ \frac{\rho_{\Lambda}}{\rho_{\Lambda}+V(\phi)} } 
\,,\label{fchoice}
\ee
where $V(\phi)$ is the scalar field potential that we want to obtain, one 
has
\be
\bar{\rho}= \frac{1}{2} \left( \frac{d\phi}{d\bar{t}} \right)^2 +V(\phi)
\,,\label{newdensity}
\ee 
which is the energy density of a new scalar field $\phi( \bar{t})$ 
obtained by changing the time variable $t\rightarrow \bar{t}(t)$. This 
transformation, which is only defined for a strictly positive $\Lambda$, 
is still implicit, as is the function $f(\rho)$ in Eq.~(\ref{fchoice}), 
which can be written down explicitly only after expressing $\phi$ as a 
function of the ``old'' density $\rho$.

For a free scalar field $\phi$ in a FLRW universe, the 
Klein-Gordon equation~(\ref{KleinGordon}) is immediately integrated to 
\be
\dot{\phi}=\frac{C}{a^3} \,, \label{firstintegral}
\ee
where $C$ is an integration constant. If the energy density is written as 
$\rho=\rho_0 /a^6 $ as usual in the fluid description, then 
$\rho_0=C^2/2$. 

The scale factor for a stiff fluid plus positive cosmological constant is 
(see, {\em e.g.}, \cite{myAJP2,Chavanis})
\be
a(t)= a_0 \left[ \sinh \left( \sqrt{3\Lambda}\, t \right) \right]^{1/3} 
\,,
\ee
where 
\be
a_0 = \left( \frac{\rho_0}{\rho_{\Lambda}} \right)^{1/6} =\left( 
\frac{C^2}{2\rho_{\Lambda}} \right)^{1/6} \,.
\ee
We can then integrate explicitly the first integral~(\ref{firstintegral}) 
obtaining
\be
\phi(t)= \frac{C}{a_0^3} \int \frac{dt}{ \sinh\left( \sqrt{3\Lambda} \, t 
\right)} = \phi_0 \ln \left[ \tanh \left( 
\frac{\sqrt{3\Lambda}}{2} \, t \right) \right] \,, \label{freescalar}
\ee
where
\be
\phi_0 = \frac{C}{\sqrt{3\Lambda} \, a_0^3} \,.
\ee
The energy density is 
\be
\rho(t) = \frac{\dot{\phi}^2}{2} = \frac{C^2}{2a_0^6} \, \frac{1}{ 
\sinh^2 \left( \sqrt{3\Lambda}\, t\right) } \,. \label{acci}
\ee
Let us express $\rho$ as  a function of the scalar field $\phi$; if this 
relation can be inverted, we will then obtain the explicit form of the 
function $f(\rho)$ that  achieves Eq.~(\ref{newdensity}). 

Exponentiating both sides of Eq.~(\ref{freescalar}) gives
\be
\mbox{e}^{\phi/\phi_0 } =\tanh \left( \frac{ \sqrt{3\Lambda} \, 
t}{2} \right)
\ee
and algebraic manipulations yield
\be
\mbox{e}^{ \sqrt{3\Lambda} \, t} = - \coth \left( \frac{\phi}{2\phi_0} 
\right) \,.
\ee
Then, it is straightforward to obtain
\be
\sinh \left(\sqrt{3\Lambda} \, t \right) = -\frac{1}{\sinh \left( 
\phi/\phi_0 \right) } \label{mahscalar}
\ee
and finally Eq.~(\ref{acci}) gives 
\be
\phi=\phi_0 \arcsinh\left( \frac{ \sqrt{2} \, a_0^3}{C} \, \sqrt{\rho} 
\right) \,,
\ee
which allows us to write explicitly the function $f(\rho)$ appearing in 
the symmetry transformation as 
\begin{eqnarray}
f(\rho) &=&  \frac{1}{ \sqrt{ 1+V/\rho_{\Lambda} } } \nonumber\\
&&\nonumber\\
&=& 
\frac{1}{  \sqrt{ 
1+ V\left( \phi_0 \arcsinh \left( \sqrt{2} \, a_0^3 
\sqrt{\rho}  /C \right)\right) / \rho_{\Lambda}  } } \,.\nonumber\\
&&
\end{eqnarray}
In terms of the comoving time $t$, we have
\be
f(t) = \frac{1}{ \sqrt{ 
1+ V\left( \phi_0 \, \ln \tanh \left( 
\frac{\sqrt{3\Lambda} \, t}{2} \right) \right)/\rho_{\Lambda} }  } \,.
\ee
The solution is not complete, though, because we need $\phi(\bar{t})$ 
instead of $\phi(t)$, which means using the first transformation 
equation~(\ref{eq:7}) to obtain $\bar{t}(t)$ and then inverting this 
relation and substituting the result into Eq.~(\ref{freescalar}). This 
means computing explicitly the integral
\be
\bar{t} (t) = \int \frac{dt}{
 \sqrt{ 1+ V \left( \phi_0 \, \ln \tanh \left( 
\frac{\sqrt{3\Lambda}\, t}{2} \right) \right)/\rho_{\Lambda}  
}} \,,\label{integral}
\ee 
and then inverting this relation to obtain $t(\bar{t})$. Unfortunately, 
these steps are usually not possible.

\section{Conclusions}
\setcounter{equation}{0}
\label{sec:5}

General symmetries of the Einstein-Friedmann equations involve 
transformations of the variables $ a, t, H, \rho$, and $P$. If $a$ or $t$ 
are transformed, $H$ cannot be transformed, and if $H$ is transformed (as 
in \cite{Chimento02}), then the only remaining choices are transforming 
$\rho$ or $P$, since $H$ is constructed out of $a$ and $t$.

Here, the term ``general'' symmetry means that no relation is imposed 
between  any two variables. For example, it is common in cosmology 
to impose the equation of state  $P=w\rho$ with $w=$~const. while 
searching for symmetries (as in 
\cite{VFSymmetry2020}), in which case one transforms the set of variables  
$ \left( \rho ,a, t,P \right) $, but the  
relation imposed between  $P$ and $\rho$ restrict the generality. 
Table~\ref{table:1} reports the possible combinations of variables 
involved in a symmetry transformation of the Einstein-Friedmann equations.

\begin{table}[]
\begin{tabular}{|c|c|c|c|}
\hline
&&&\\ 
~$\rho ,a,t,P$~ & ~$\rho ,a,t$~ & ~$\rho ,a,P$~ & ~$\rho ,t,P$~ \\ 
&&&\\  \hline
&&&\\ 
$a,t,P$        & $H,\rho,P$  & $\rho ,a $  & $\rho,t$   \\ 
&&&\\ \hline
&&&\\ 
$\rho,P$      & $a,t$        & $a,P$        & $t,P$        \\ 
&&&\\ \hline
&&&\\ 
$H,\rho$      & $H,P$         & --          & --           \\ 
&&&\\ \hline
\end{tabular}
\caption{The possible combinations of variables 
involved in a symmetry transformation of the Einstein-Friedmann 
equations.}  \label{table:1}
\end{table}

The $\left( H, \rho ,P\right)$ case has already been partially covered by 
Chimento \cite{Chimento02}.  In this work, we have studied an extension of 
his symmetry transformations for $\Lambda >0$. Any symmetry transforming 
the scale factor cannot be made more general, because it demands a 
restriction of variables (these symmetries are still of interest when the 
barotropic, linear, and constant equation of state $P=w\rho$ is imposed).

The possible general symmetries transforming only two variables include 
$\left( \rho ,t \right), \left(\rho,P \right), \left( t,P \right), \left( 
H, \rho \right)$, and $\left( H,P \right)$. However, it can be easily 
shown that these are not suitable for generating general symmetries. Only 
one possible general symmetry remains, {\em i.e.}, the one involving 
$\left( \rho, t,P \right)$ and, to the best of our knowledge, this was  
not discussed in the literature yet (however, if one does not touch $a$, 
rescaling the time $t$ is equivalent to rescaling $H$ together with $\rho$ 
and $P$, which was done in \cite{Chimento02} for vanishing cosmological 
constant).

The situation in which the new symmetry reported here would be most useful 
is in the generation of new scalar field solutions for a specified 
potential $V(\phi)$ from the corresponding solution for a free scalar 
field 
(plus positive $\Lambda$). Unfortunately, although the generating method 
exposed 
in Sec.~\ref{sec:4} carries through almost to the end, two obstacles will, 
in general, forbid one to obtain results. First, one needs to compute 
explicitly the integral~(\ref{integral}) in terms of elementary functions, 
which in 
general is not possible. Second, one needs to invert analytically the 
relation $\bar{t}(t)$ thus obtained, which is also, by all means, not 
guaranteed. It is interesting, however, that a tall order such as 
generating solutions of the non-linear Einstein-Friedmann-Klein-Gordon 
equations from a free field solution can be carried through to such an 
extent. In the case $\Lambda=0$ studied in \cite{Chimento02}, one cannot 
generate  a potential $V(\phi)$ as we did here and the free scalar field 
remains free after the transformation is performed.

Overall, the conclusion emerging is that the general transformations of the 
kind~(\ref{eq:7})-(\ref{eq:9}) (or their restriction to $\Lambda=0$ 
studied in \cite{Chimento02}) are not very useful in generating new 
inflationary solutions for the early universe as one would have hoped.

\begin{acknowledgments} 

This work is supported, in part, by the Natural Sciences \& Engineering 
Research Council of Canada (Grant no. 2016-03803 to V.F.) and by Bishop's 
University.

\end{acknowledgments}

\appendix
\section{Explicit examples of symmetry transformations}
\renewcommand{\theequation}{A.\arabic{equation}}

Here we show other explicit examples of transformations of the 
kind~(\ref{eq:7})-(\ref{eq:9}) that carry a FLRW universe with a perfect 
fluid and $\Lambda>0$ into one with a fluid with non-linear equations of 
state plus $\Lambda$.

For perfect fluids with constant equation of state $P=w\rho$, the 
transformed equation of state is 
\be
\bar{P }=-\bar{\rho}+(w+1)  
\frac{\left[ f-2\left( \rho+\rho_{\Lambda} \right) f'\right]}{ 
 f^3} \, \rho \,;
\ee
this expression neds to be rewritten in terms of barred variables only. 
To this end, one can search for a function $f(\rho)$ such that
\be
\frac{\left[ f-2\left(\rho+ \rho_{\Lambda} \right) f'\right]}{ 
 f^3}\rho=g\left( \bar{\rho}\right) =g\left(\frac{\left(1-f^2\right)}{ f^2} 
\rho_{\Lambda} + \frac{\rho}{f^2}\right) \,. \label{ODE2}
\ee
Certain functions $g(\bar{\rho})$ allow the above ordinary differential 
equation to be solved for interesting  equations of state of the form 
\be
 \bar{P}=\left[\frac{  g\left( \bar{\rho} 
\right) }{\bar{\rho}}(w+1)-1\right]\bar{\rho} \,.
\ee

Solving Eq.~(\ref{ODE2}) with $g\left(\bar{\rho}\right)=\alpha\bar{\rho}$ 
we obtain
\be
f(\rho)=\sqrt{ \frac { \left( \rho+ \rho_{\Lambda} \right) }{ \gamma
\rho^{\alpha}+\rho_{\Lambda} }} \,,
\ee
which brings  a perfect fluid with constant equation of state into another 
one, and has been studied in Sec.~\ref{sec:3}.

Another example is given by solving Eq.~(\ref{ODE2}) with $g\left( 
\bar{\rho} \right) =\alpha  \tanh \bar{\rho}$, which gives 
\be 
f(\rho)=\sqrt {\frac{\rho+\rho_{\Lambda} }{
\arcsinh\left( \gamma \rho^{\alpha} \right)+\rho_{\Lambda}} }\,,
\ee
where $\alpha$ and $\gamma$ are constants, with $\alpha$ dimensionless 
and $\left[ \gamma \right] = \left[ \rho^{-\alpha} \right]$. Then, one 
obtains   
\be
\bar{\rho}=\arcsinh \left( \gamma \, \rho^{\alpha} \right)
\ee
and
\be
\bar{P}=  \alpha  (w+1) \tanh \bar{\rho} - \bar{\rho} \,.
\ee

The third example is generated by solving Eq.~(\ref{ODE2}) with $g\left( 
\bar{\rho} \right)=\alpha \, \mbox{e}^{\beta(\bar{\rho})}$. This solution 
generates
\be
f(\rho)=\sqrt{ -\frac{\beta\, \left( \rho+ \rho_{\Lambda} \right) 
}{ \ln  \left[ \alpha \beta  \left( \gamma-\ln (\rho/\rho_0)  \right)  
\right] 
-\beta \rho_{\Lambda}  } } \,,
\ee
with $\alpha, \beta$, and $\gamma$ dimensionless constants, while $\rho_0$ 
is a constant with the dimensions of an energy density. 
This choice gives
\be
\bar{\rho}= - \frac{ \ln\left\{ \alpha \beta \left[ 
\gamma-\ln (\rho/\rho_0)    \right]  \right\} }{ \beta }
\ee
and
\be
\bar{P}= \alpha (w+1) \mbox{e}^{\beta \, \bar{\rho}}  
-\bar{\rho} \,.
\ee

A fourth example is given by the solution 
$g(\bar{\rho})=\alpha/ \bar{\rho}$ of Eq.~(\ref{ODE2}), which 
yields
\be
f(\rho)=\sqrt{ \frac{\rho+\rho_{\Lambda} }{ \sqrt {\gamma+2\alpha  \ln 
( \rho/\rho_0)  } + \rho_{\Lambda} }  } \,,
\ee
and 
\be
\bar{\rho}= \sqrt{\gamma+2\alpha  \ln ( \rho/\rho_0) } \,,
\ee
\be
\bar{P}=  \frac{\alpha \left(w+1\right)}{\bar{\rho} }- \bar{\rho} \,.
\ee

Our last example is generated by $g(\bar{\rho})=\alpha \, {\bar{\rho}}^2$ 
and 
the subsequent
\begin{widetext}
\be
f(\rho) ={\frac {\sqrt {\alpha \left( \ln(\rho/\rho_0) + \gamma 
\right)\left( 
\alpha \rho_{\Lambda}\ln(\rho/\rho_0) +\alpha\gamma \rho_{\Lambda} 
+1 \right) \left(\rho+\rho_{\Lambda} \right) }}{\alpha
\left( \ln(\rho/\rho_0)  +\gamma \right) \rho_{\Lambda}} }\,,
\ee
\end{widetext}
which gives
\be
\bar{\rho}=-{\frac 
{\rho_{\Lambda} }{\alpha  (\ln(\rho/\rho_0) + \gamma)\rho_{\Lambda} + 1}}
\ee
and
\be
\bar{P}=\left[\alpha\bar{\rho}(w+1)-1\right]\bar{\rho} \,.
\ee

Rather cumbersome symmetry mappings, that we do not report, are generated 
by the solutions $g\left( \bar{\rho}\right) = \alpha \bar{\rho}^3 , \,  
\alpha 
\bar{\rho}^4 $ of Eq.~(\ref{ODE2}). For powers of $\bar{\rho}$ higher 
than four, the relevant expressions become very long and involved.


\begin{thebibliography}{99}

\bibitem{Wald} R.M. Wald, {\em General Relativity} (Chicago University 
Press, Chicago, 1984).

\bibitem{Chimento02} L.P. Chimento, 
{\em Phys. Rev. D} {\bf  65}, 063517 (2002). 

\bibitem{SD2} J.M. Aguirregabiria, L.P. Chimento, A. Jakubi, and R. 
Lazkoz, 
{\em Phys. Rev. D} {\bf 67}, 083518 (2003). 

\bibitem{SD3} L.P. Chimento and R. Lazkoz,
{\em  Phys. Rev. Lett.} {\bf 91}, 211301 (2003). 

\bibitem{SD4} M.P. Dabrowski, T. Stachowiak, and M. Szydlowski,  
{\em Phys. Rev. D} {\bf 68}, 103519 (2003).

\bibitem{SD5} J.M. Aguirragabiria, L.P. Chimento, and R. Lazkoz, 
{\em Phys. Rev. D} {\bf 70}, 023509 (2004).

\bibitem{SD6} G. Calcagni,
{\em Phys. Rev. D} {\bf 71}, 023511 (2005).

\bibitem{SD7}  M. Szydlowski, W. Godlowski, and R. Wojtak,
{\em Gen. Relativ. Gravit.} {\bf 38}, 795 (2006).

\bibitem{SD8} L.P. Chimento and R. Lazkoz, 
{\em Classical Quantum Gravity} {\bf 23}, 3195 (2006).

\bibitem{SD9} L.P.  Chimento and W. Zimdhal,
{\em Int. J. Mod. Phys. D} {\bf 17}, 2229 (2008).

\bibitem{SD10} L.P. Chimento and D. Pavon, 
{\em  Phys. Rev. D} {\bf 73}, 063511 (2006).

\bibitem{SD11} M.P. Dabrowski, C. Kiefer, and B. Sandhoefer, 
{\em Phys. Rev. D} {\bf 74}, 044022 (2006).

\bibitem{SD12} Y.-F. Cai, H. Li, Y.-S. Piao, and X. Zhang, 
{\em Phys. Lett. B} {\bf 646}, 141 (2007). 

\bibitem{SD13} L.P. Chimento, F.P. Devecchi, M.I. Forte, and G.M. Kremer, 
{\em Classical Quantum Gravity} {\bf 25}, 085007 (2008).

\bibitem{SD14} M. Cataldo and L.P. Chimento, 
{\em Int. J. Mod. Phys. D} {\bf 17}, 1981 (2008).

\bibitem{SD15} S. Capozziello, E. Piedipalumbo, C. Rubano, and P. 
Scudellaro, 
{\em Phys. Rev. D} {\bf 80}, 104030 (2009).

\bibitem{SD16} J. Wang, T. Lan, and S.-P. Yang, 
{\em J. Theor. Phys.} {\bf 1}, 62 (2012).

\bibitem{SD17} S. Capozziello and V. Faraoni, {\em Beyond Einstein 
Gravity} (Springer, New York, 2010).

\bibitem{SD18}  Y.-F. Cai, E.N. Saridakis, M.R. Setare, and J.-Q. Xia, 
{\em Phys. Rep.} {\bf 493}, 1 (2010).

\bibitem{SD19} L. Pucheu and M. Bellini,
{\em Nuovo Cimento B} {\bf 125}, 851 (2010)  

\bibitem{SD20} L.P. Chimento, R. Lazkoz, and M.G. Richarte, 
{\em Phys. Rev. D} {\bf 83}, 063505 (2011).

\bibitem{SD21} V. Faraoni,
{\em Phys. Lett. B} {\bf 703}, 228 (2011).

\bibitem{susy1} H.C. Rosu and K.V. Khmelnytskaya,
{\em SIGMA} {\bf 7}, 013 (2011).

\bibitem{susy2} H.C. Rosu and P. Ojeda-May, 
{\em Int. J. Theor. Phys.} {\bf 45}, 1191 (2006).

\bibitem{susy3} M. Nowakoswki and H.C. Rosu,  
{\em Phys. Rev. E} {\bf 65}, 047602 (2002). 

\bibitem{Pailasetal20} T. Pailas, N. Dimakis, A. Paliathanasis, P.A. Terzis, and T. Christodoulakis, 
arXiv:2005.11726 [gr-qc].

\bibitem{Lieolder} 
T. Christodoulakis, A. Karagiorgos, and A. Zampeli, {\em Symmetry} {\bf 10}, 70 (2018);
M. Tsamparlis and A. Paliathanasis, {\em Symmetry} {\bf 10}, 233 (2018);
G. Gionti and A. Paliathanasis, {\em Mod. Phys. Lett. A} {\bf 33}, 1850093 (2018);
A. Paliathanasis, M. Tsamparlis, S. Basilakos, and J.D. Barrow, {\em Phys. Rev. D} {\bf 93}, 043528 
(2016);
A. Paliathanasis and S. Capozziello, {\em Mod. Phys. Lett. A} {\bf 31}, 1650183 (2016);
A. Zampeli, T. Pailas, P.A. Terzis, and T. Christodoulakis, {\em J. Cosmol. Astropart. Phys} {\bf 16}, 
066 (2016);
T. Christodoulakis, N. Dimakis, P.A. Terzis, B. Vakili, E. Melas, and T. Grammenos, {\em Phys. Rev. D}
{\bf 89}, 044031 (2014);
A. Paliathanasis and M. Tsamparlis, {\em Phys. Rev. D} {\bf 90}, 043529 (2014);
T. Christodoulakis, N. Dimakis, and P.A. Terzis, {\em J. Phys. A: Math. Theor.} {\bf 47}, 095202 
(2014);
N. Dimakis, T. Christodoulakis, and P.A. Terzis, {\em J. Geom. Phys.} {\bf 77}, 97 
(2012); 
S. Basilakos, M. Tsamparlis, and A. Paliathanasis, {\em Phys. Rev. D} {\bf 83}, 103512 (2011);
A. Paliathanasis, M. Tsamparlis, and S. Basilakos, {\em Phys. Rev. D} {\bf 84}, 123514 (2011).

\bibitem{Borowiec14} A. Borowiec, S.  Capozziello, M. De Laurentis, F.S.N. 
Lobo, A. Paliathanasis, M. Paolella, and A. Wojnar, {\em Phys. Rev. D} 
{\bf 91}, 023517 (2015).

\bibitem{Mariafelicia15} S. Capozziello, M. De Laurentis, and R. 
Myrzakulov, {\em Int. J. Geom. Meth. Mod. Phys.} {\bf 12}, 1550095 (2015).

\bibitem{Mariafelicia16} S. Capozziello, M. De Laurentis, K.F. 
Dialektopoulos, {\em Eur. Phys. J. C} {\bf 76}, 629 (2016).
 
\bibitem{Mariafelicia20} E. Piedipalumbo, M. De Laurentis, and S. 
Capozziello
{\em Phys. Dark Univ.} {\bf 27} 100444 (2020).

\bibitem{VFSymmetry2020} V. Faraoni, {\em Symmetry} {\bf 12}, 147 (2020). 	

\bibitem{Madsen1}  M.S. Madsen, {\em Classical Quantum Gravity} {\bf 5}, 
627 
(1988).

\bibitem{Madsen2} M.S. Madsen, {\em Astrophys. Space Sci.} {\bf 113}, 205 
(1985).

\bibitem{Pimentel} L.O. Pimentel, {\em Classical Quantum Gravity} {\bf 
6}, L263 (1989).

\bibitem{Faraoni12} V. Faraoni, {\em Phys. Rev. D} {\bf 85}, 024040 
(2012).

\bibitem{Semiz12} I. Semiz, {\em Phys. Rev. D} {\bf 85}, 068501 (2012).

\bibitem{Ballesteros} G. Ballesteros, D. Comelli, and L. Pilo, {\it Phys. 
Rev. 
D} {\bf 94}, 025034 (2016).

\bibitem{Jeremy} V. Faraoni and J. C\^ot\'e, {\em Phys. Rev. D} {\bf 98}, 
084019 (2018).

\bibitem{myAJP2} V. Faraoni, {\em Am. J. Phys.} {\bf 69}, 372 (2001).

\bibitem{BayinCooperstockFaraoni} S.\c{S}. Bayin, F.I. Cooperstock, and 
V. Faraoni, {\em Astrophys. J.} {\bf 428}, 439 (1994).

\bibitem{Chavanis} P.H. Chavanis, {\em Phys. Rev. D} {\bf 92}, 103004 
(2015).

\end{thebibliography}
\end{document}